\documentclass[]{iopart}
\usepackage{iopams}
\usepackage[]{graphicx}
\usepackage{color}

\newcommand{\colr}{}

\newcommand{\IF}{\vartheta}

\newcommand{\E}{\textrm{E}}
\newcommand{\fL}{{\frak L}}

\begin{document}

\title[]{Bayesian inference of a negative quantity from positive measurement results}
\author{D Calonico, F Levi, L Lorini and G Mana}
\address{INRIM - Istituto Nazionale di Ricerca Metrologica, Str.\ delle Cacce 91, 10135 Torino, Italy}
\ead{g.mana@inrim.it}

\begin{abstract}
In this paper the Bayesian analysis is applied to assign a probability density to the value of a quantity having a definite sign. This analysis is logically consistent with the results, positive or negative, of repeated measurements.
Results are used to estimate the atom density shift in a caesium fountain clock. {\colr The comparison with the classical statistical analysis is also reported and the advantages of the Bayesian approach for the realization of the time unit are discussed.}
\end{abstract}

\submitto{Metrologia}
\pacs{02.50.Cw, 02.50.Tt, 06.20.Dk, 07.05.Kf}
% 02.50.Cw Probability theory
% 02.50.Tt Inference methods
% 06.20.Dk Measurement and error theory
% 07.05.Kf Data analysis: algorithms and implementation; data management

%\baselineskip 8mm

\section{Introduction}
{\colr The Bayesian analysis is acquiring increasing importance in metrology; a tutorial guide can be found in Sivia's book \cite{sivia} and a comprehensive and authoritative account is the treatise by Jaynes \cite{jaynes}. The present paper illustrates the use of the Bayes theorem to infer the collisional coefficient of atom-fountain clocks -- from theory, a negative one -- from the results of a sequence of measurements.} The paper illustrates in a very simple way the differences between the orthodox and the Bayesian analyses, which differences are usually unnoticed.

When the data are used to estimate the measurand, the orthodox analysis concerns the probability density of the estimate \cite{gum}. Instead, the Bayesian approach concerns the probability density of the measurand, given the single data set generated by measurement and prior possible information. If the data are unbiased normal variables, in the absence of prior information, the same multivariate Gaussian function describing the estimate distribution is also the probability density of the measurand. Since this function is invariant when the estimate is exchanged for the measurand, the orthodox and the Bayesian inferences, though conceptually different, are numerically the same. This explains the limited awareness of the risk of using the solution of one problem as the solution of the other.

\begin{figure}\centering
\includegraphics[width=7cm]{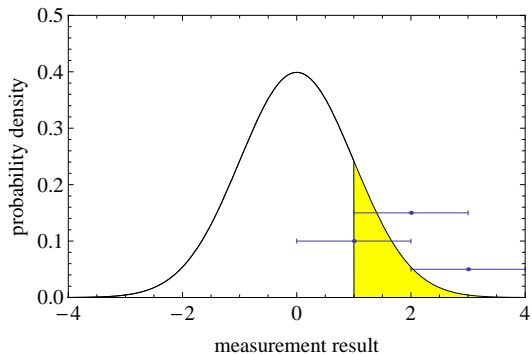}
\caption{A sample from a Gaussian distribution with zero mean and unit variance has a 50\% probability to be positive. If it lies in the 16\% distribution tail (filled in yellow), the classical confidence interval is in the positive region. Examples of measurement results with an associated uncertainty entirely in the positive region are also shown.} \label{f00}
\end{figure}

Had the data been sampled from a distribution not having the above symmetry, the difference between the probability densities of the estimate and of the measurand cannot be unnoticed. An interesting case is that of a positive measurement result when the measurand is negative. Though the measurement result is an unbiased estimate, this inference looks quite strange. {\colr As exemplified in Fig.\ \ref{f00}, if the measurand approaches zero when compared to the measurement uncertainty, there is nothing unusual about obtaining a positive value. Assuming a Gaussian sampling distribution, there is almost a 50\% probability that the measurement result is positive and about a 16\% probability that the 68\% classical confidence interval (the result plus/minus the uncertainty) lies in the non-physical region. Though the confidence interval is instrumental only in the assessment of the estimate and not of the measurand \cite{Neyman}, this view is not shared by most metrologists. They figure out a Gaussian function centred on the estimated value and discuss whether the measurand lies within the confidence interval. In the above situation, the consequent paradoxical confidence interval in the non-physical region is explained by observing that, according to a seminal 1937 paper by Neyman \cite{Neyman}, the confidence interval is so defined as to keep strictly to a statement about the probability density of the measurement result without considering the probability density of the measurand. According this statement, independently of the measurand value, only 68\% of the intervals obtained from a set of measurements will contain the measurand.}

\section{Repeated measurements of a positive quantity}
\subsection{statement of the problem}
Let us suppose that the measurement of a quantity $y$, known to be negative, yields value $y_1$. The measurement result $y_m$ is assumed normally distributed about the value $y_0$ of the unknown quantity, with a known standard deviation $\sigma_1$. The probability density of obtaining the measurement value $y_1$ is
\begin{equation}\label{gauss}
 P_{y_m}(y_m=y_1|y=y_0) = \frac{1}{\sqrt{2\pi}\;\sigma_1} \exp\bigg[ \frac{-(y_1-y_0)^2}{2\sigma_1^2} \bigg] .
\end{equation}
The question is how to account for $y_0<0$ and to infer a consistent value for the measurand.

\subsection{Bayesian solution}
In the absence of any additional information, the Bayesian considers the $y_0<0$ constraint in the probability density of the measurand values before the measurement result is know,
\begin{equation}\label{prior}
 P_y(y=y_0) = \IF(-y_0) ,
\end{equation}
where $\IF({x})$ is the heaviside function. According to the Bayes theorem, the post-data probability density of the possible $y$ values, given the measurement result $y_m=y_1$, is the tail of $P_{y_m}(y_m=y_1|y=y_0)$ lying in the $y_0<0$ interval,
\begin{eqnarray}\label{post}
 P_y(y=y_0|y_m=y_1) &= &\frac{P_{y_m}(y_m=y_1|y=y_0)P_y(y=y_0)}{P_{y_m}(y_m=y_1)} \nonumber \\
 &= &\frac{2\exp\big[-(y_0-y_1)^2/2\big] \IF(-y_0)}{\sqrt{2\pi}\,\textrm{erfc}(y_1/\sqrt{2})} ,
\end{eqnarray}
where we set $\sigma_1=1$ since this requires only a trivial change of variables. The probability density (\ref{post}) embeds both the information available independently of the measurement result, that is, $y_0<0$, and the information delivered by the $y_1$ datum.

\subsection{Bayesian inferences}
{\colr To convert the measurand post-data probability density (\ref{post}) into a single numerical estimate, $\hat y$, the loss, $\fL(\hat y-y)$, associated with the estimate error must be specified. The $\fL(\hat y-y)$ function maps the error onto a number representing the cost associated with a wrong estimate. The optimal estimator minimizes the expected loss over the measurand distribution, that is, $\hat y = \arg_{\hat y} \min \E\big[ \fL(\hat y-y) \big]$, where
\begin{equation}
 \E\big[ \fL(\hat y - y) \big] = \int_{-\infty}^{+\infty} \fL(\hat y - y_0)P_y(y_0|y_1)\rmd y_0 .
\end{equation}
The squared-error, that is, $\fL(\hat y-y)=(\hat y-y)^2$, indicates the mean \cite{math},}
\numparts\begin{equation}\label{mean}
 \hat y = y_1 - \frac{\sqrt{2/\pi}\exp(-y_1^2/2)}{\textrm{erfc}(y_1/\sqrt{2})} ,
\end{equation}
the absolute-difference, that is, $\fL(\hat y-y)=|\hat y-y|$, indicates the median, which is implicity given by
\begin{equation}\label{median}
 -2\textrm{erf}\big[ (\hat y - y_1)/\sqrt{2} \big] = 1 + \textrm{erf}(y_1/\sqrt{2}) ,
\end{equation}
and $\fL(\hat y-y)=$ const. indicates the mode,
\begin{equation}\label{mode}
 \hat y = \left\{ \begin{array}{ll}
                    y_1 & \textrm{if}\; y_1<0 \\
                    0   & \textrm{if}\; y_1\ge 0
                  \end{array} \right. .
\end{equation}\endnumparts
Confidence intervals are easily expressed by integrating (\ref{post}) to obtain the relevant cumulative distribution function.

\begin{figure}\centering
\includegraphics[width=7cm]{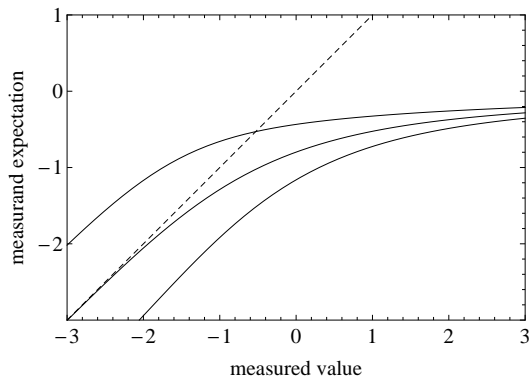}
\caption{Expected value of a negative quantity given a measurement value sampled from an unbiased Gaussian distribution with unit variance. Lower and upper lines indicate the plus/minus one standard deviation interval. The dashed line is the orthodox estimate of the measurand.} \label{f01}
\end{figure}

\begin{figure}\centering
\includegraphics[width=7cm]{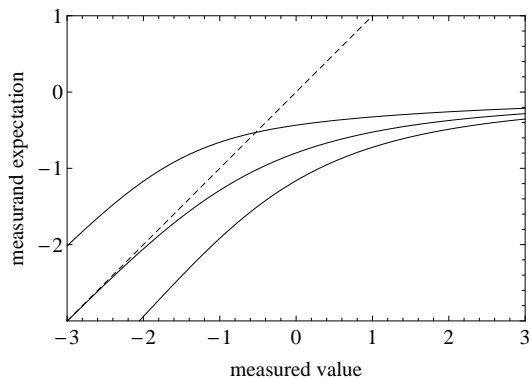}
\caption{Median of a negative quantity given a measurement value sampled from an unbiased Gaussian distribution with unit variance. Lower and upper lines are the first and third quartile. The dashed line is the orthodox estimate of the measurand.} \label{f02}
\end{figure}

The expected value and the median of the measurand are shown in Figs. \ref{f01} and \ref{f02} together with the standard deviation and quartiles. When the measurement result is negative and large, both the expected value and the median coincide with the measured value. In this case, both the orthodox and Bayesian analyses deliver the same estimate. In fact, $P_y(y=y_0|y_m=y_1)$ is identical with $P_{y_m}(y_m=y_1|y=y_0)$ for all practical purposes and the symmetry $P_{y_m}(y_m=y_1|y=y_0)=P_{y_m}(y_m=y_0|y=y_1)$ allows the two probability densities, though conceptually different, to be identified. When the measurement result approaches zero, or even is positive, the difference between the orthodox and the Bayesian analyses is evident. The measurement result is now a very poor estimate of the measurand value. On the contrary, as shown in Figs. \ref{f01} and \ref{f02}, the Bayesian approach, based on the post-data probability density, does not have any difficulty. {\colr The smaller (in absolute terms) the measurement result is, the nearer to zero, but negative, is the estimate; it is to be noted that, in this case, also the uncertainty of the estimate decreases.}

\subsection{extension to repeated measurements}
The previous analysis can be extended to $N$ repeated independent measurements. In this case, the post-data probability density of the $y$ values is
\begin{eqnarray}\label{post-multi}\fl
 P_y(y=y_0|y_m=\{y_1,y_2,\;...\;y_N\}) &= &\frac{P_y(y=y_0) \prod_{i=1}^N P_{y_m}(y_m=y_i|y=y_0)}
 {P_{y_m}(y_m=\{y_1,y_2,\;...\;y_N\})} \nonumber \\
 &= &\frac{2\exp\big[-(y_0-\bar y)^2/2\big] \IF(-y_0)}{\sqrt{2\pi}\,\big[1+\textrm{erf}(\bar y/\sqrt{2})\big]} ,
\end{eqnarray}
where $\bar y$ is the weighed sample mean and the $\bar y$ variance, $\sigma_{\bar y}^2=\big( \sum 1/\sigma_i^2 \big)^{-1}$, has been set to one. To compress (\ref{post-multi}) into a single numerical estimate of $y$, we can still use (\ref{mean}-$c$), where the sample mean $\bar y$ substitutes for $y_1$.

In another way to process data, we observe that $\bar y$ is an estimate of $y$ normally distributed about $y_0$. If the data are numerous, $\bar y$ is expected to be negative. Therefore, it might seen to be an optimal estimate. However, from a Bayesian viewpoint, given $\bar y$ and $y_0>0$, the starting points to calculate the probability density of the measurand values are again (\ref{gauss}) and (\ref{prior}), where $\bar y$ and $\sigma_{\bar y}$ substitute for $y_1$ and $\sigma_1$. The probability density of the measurand is identical to (\ref{post-multi}), as expected. Hence, if the cost function is quadratic, the optimal estimate is the expected value of the measurand over the probability distribution of its possible values, given by (\ref{mean}). But this expected value is greater than $\bar y$, as shown in Fig.\ \ref{f01}. To explain why the Bayesian estimate (\ref{mean}) differs from the orthodox, we must observe that now, owing to the prior information, the probability densities of the estimate and of the measurand are not the same function.

\section{Evaluation of the collisional coefficient of a Cs fountain}
An interesting application of previous results is offered by primary frequency metrology. Ultracold Cs fountains realize the time unit, the second, with a relative uncertainty of few parts in $10^{-16}$. This achievement has been made possible by the use of atomic Cs samples cooled down to about 1 $\mu$K. A detailed description of this frequency standard can be found in \cite{M06} and references therein. Caesium fountains implement a Ramsey spectroscopy scheme, where a sample of cold atoms is launched in a ballistic flight. During the flight, the atoms go upwards and then come back downwards passing twice through a microwave cavity and interacting with a microwave radiation at 9 192 631 770 Hz. Then, an optical detector counts how many atoms have undergone the hyperfine transition induced by the microwave radiation. The transition probability is then calculated, thus obtaining the frequency difference between the Cs reference and the interrogating microwave. The SI second, the definition of which dates from the 13th Conf\'erence G\'en\'erale des Poids et Mesures in 1967, is the duration of 9 192 631 770 periods of the radiation corresponding to the transition between the two hyperfine levels of the ground state of the Cesium 133 atom. Therefore, the realization of the second should rely on a single and unperturbed atom, while atomic fountains use samples of about $10^{5}-10^{6}$ atoms.

In these samples, collisions occur among the ultra-cold atoms with consequent perturbation of the energy levels and shift of the atomic reference frequency. This shift, called collisional or atom density shift, is proportional to the atom density and it must be carefully evaluated in order to correct the frequency standard. Its uncertainty is one of the main contributions to the accuracy budget of an atomic fountain. As collisions play a crucial role in high precision clocks, they have been the subject of many theoretical and experimental studies \cite{Williams}.

It has been demonstrated theoretically and experimentally \cite{Winands} that, if a fountain is operated in a way to make the collision energy high during the whole ballistic flight, the collisional shift has a weak dependance on the atom temperature. This is the case, for example, of direct molasses capture or molasses expansion after magneto-optical-trap capture \cite{M06}. Moreover, if the atomic sample is a quantum mixture of the hyperfine eigenstates $|F=3, m_F=0 \rangle$ and $|F=4, m_F=0 \rangle$, as occurs in the fountain operation, the collisional shift has also a negligible dependence on the mixture ratio and it is linearly dependent on density through a negative coefficient.

Provided some technical parameters are known and kept stable the density shift estimation is possible, but with high uncertainty. An evaluation of the density shift of lower uncertainty is possible by a differential measurement technique: the fountain is operated alternately at high and at low atom densities. Then, the two measurement results are used to extrapolate the frequency to the zero density, with use of a linear regression. This differential measurement evaluates the atom density frequency sensitivity; we refer to this sensitivity as the collisional coefficient. This measurement technique requires that the parameters linking the output signal to the atom density remain stable only for the differential measurement time instead of during the whole measurement duration. The evaluation of the density shift must be repeated regularly to cope with the possible instabilities.

Since the leverage of the differential measurement is not particularly high, the frequency shift often approaches the clock resolution. Therefore, the measured values of the collisional coefficient have an uncertainty comparable to their magnitude. Table 1 records measurement results, together with their uncertainties, collected during an accuracy evaluation of INRIM fountain IT-CsF1. The collisional coefficient value expected from theory is negative and the overall measurement values confirm this assertion. Nevertheless, some of the values are positive, in agreement with a Gaussian dispersion. These data are suitable for Bayesian analysis; in fact, when using an orthodox approach, we are not able to take into account that we know that the collisional coefficient is negative.

\begin{table}
 \centering
 \caption{Clock-frequency sensitivity to the atom density (collisional coefficient) as measured in a Cs fountain accuracy evaluation}\label{T1} \vspace{2mm}
 \begin{tabular}{rrrrrrrrrr}
   \hline\hline
   \multicolumn{5}{c}{collisional coefficient (uncertainties) / arbitrary units}\\
   \hline
$-2.97 (5.19)$ &$ -3.27(3.96)$ &$ -6.41(3.64)$ &$ -1.42(4.85)$ &$ +4.31(3.76)$ \\
$ -1.88(3.53)$ &$ -5.35(3.91)$ &$ -5.71(3.69) $ &$ -4.04(3.49)$ &$ -7.48(3.66)$  \\
$-1.37(3.71)$ &$ -2.41(3.66)$ &$ -3.60(3.93)$ &$ -4.79(4.54)$ &$ -4.56(4.43)$ \\
$ -1.93(4.56)$ &$ -6.41(3.57)$ &$ -2.23(3.48)$ &$ -4.75(3.72)$ &$ -1.47(3.49)$  \\
$-4.56(3.29)$ &$ +2.59(3.71)$ &$ -2.76(3.75)$ &$ +0.47(3.72)$ &$ -0.97(3.98)$ \\
$ +2.17(3.95)$ &$ -4.75(3.98)$ &$ +0.58(3.82)$ &$ -0.81(3.79)$ &$ -3.64(4.22)$  \\
$-5.71(4.61)$ &$ +2.99(4.89)$ &$ +0.97(6.82)$ &$ -2.90(4.82)$ &$ +2.17( 4.66)$ \\
$ -5.53( 4.01)$ &$ -0.40(4.22)$ &$ -4.48(3.81)$ &$ +0.10(3.84)$ &$ -3.87(3.93)$  \\
$-3.43(4.32)$ &$ -5.13(4.33)$ &$ -5.42(4.00)$ &$ -3.19( 4.22)$ &$ -1.35( 4.40)$ \\
$ -0.80( 4.32)$ &$ -0.08(4.28)$ \\
   \hline
   \end{tabular}
\end{table}

\begin{figure}\centering
\includegraphics[width=7cm]{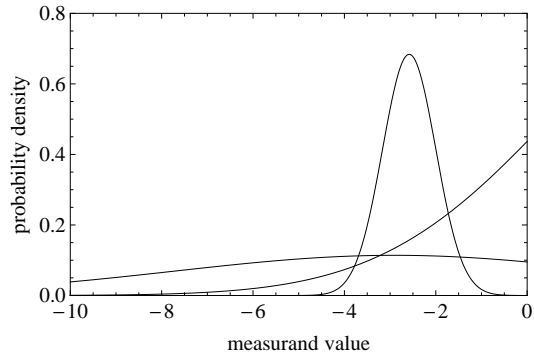}
\caption{Post-data probability density of the collisional coefficient, given the measurement values and uncertainty in Table 1 (Gaussian peak) and the measurement values $4.31 \pm 3.76$ (exponential line) and $-2.90 \pm 4.82$ (broad Gaussian curve).} \label{f03}
\end{figure}

The post-data probability density of the measurand is shown in Fig. \ref{f03}; owing to the relatively sharp peak, the Bayesian cut of the positive tail of the distribution has no practical effect. {\colr Both the weighed mean of the sample} and the Bayesian expected value of the measurand are $-2.58 \pm 0.58$. We can also infer the sensitivity value from the $4.31 \pm 3.76$ datum, the confidence interval of which is entirely in the non-physical region. The result is $-1.45_{-1.60}^{+0.96}$ (median plus/minus quartiles). Despite the unsatisfactory datum, the Bayesian inference is really good, when compared to the much more accurate value inferred from many measurement repetitions. In general, when the datum is negative the Bayesian and orthodox estimates tend to coincide. The worst case is when the measurement result is negative, as expected, but has large uncertainty; for example, if the $-2.90 \pm 4.82$ datum is considered, the Bayesian median plus/minus quartiles is  $-4.59_{-3.40}^{+2.68}$.

\section{Conclusions}
The Bayes theorem accounts for prior information at the very beginning of data analysis, in a way ensuring the logical consistency of statistical inferences. In the specific example here considered, the theorem demonstrates capable to cope with a problem the orthodox solution of which is unsatisfactory.

As regards primary frequency metrology, the results we obtained imply two main consequences. Firstly, by use of the sign information, the density shift of a Cs-fountain primary standard can be estimated within a given uncertainty from a smaller data set. Hence, the number of collisional coefficient measurements can be reduced and the density shift evaluation can be made faster, without increasing the uncertainty. Besides, the reduction of the evaluation time is a desirable feature of a Cs fountain. Secondly, each measured density coefficient is immediately used to extrapolate the fountain frequency to zero density. Therefore, a proper use of the sign information to extrapolate a single measurement pair to zero density will improve the measurement capabilities and the detection of failures and anomalies. This will be the subject matter of future work.


\section*{References}

\begin{thebibliography}{99}
\bibitem{sivia}
 Sivia D S and Skilling J 2007 {\it Data Analysis: a Bayesian Tutorial} (Oxford: Oxford University Press)
\bibitem{jaynes}
 Jaynes E T 2003 {\it Probability Theory: the Logic of Science} (Cambridge: Cambridge University Press)
\bibitem{gum}
 BIPM, IEC, IFCC, ILAC, ISO, IUPAC, IUPAP and OIML 1995 {\it Guide to the expression of uncertainty in measurement} (Geneva, Swizerland: International
Organization for Standardization)
\bibitem{Neyman}
 Neyman J 1937 Outline of a theory of statistical estimation based on the classical theory of probability {\it Philos.\ Tran.\ R.\ Soc.\ London A}
{\bf 236} 333-80
\bibitem{math}
 The calculations have been made with Mathematica, Wolfram Research Inc.\
\bibitem{M06}
 Levi F, Calonico D, Lorini L and Godone A 2006 IEN-CsF1 primary frequency standard at INRIM: accuracy evaluation and TAI calibrations {\it Metrologia} {\bf 43} 545-555
\bibitem{Williams}
 Leo P J, Julienne P S, Mies F H and Williams C J 2001 Collisional Frequency Shifts in 133Cs Fountain Clocks {\it Phys.\ Rev.\ Lett.} {\bf 86} 3743-3746
\bibitem{Winands}
 Szymaniec K, Chalupczak W, Tiesinga E, Williams CJ, Weyers S and Wynands R 2007 Cancellation of the Collisional Frequency Shift in Caesium Fountain Clocks {\it Phys.\ Rev.\ Lett.} {\bf 98}, 153002
\end{thebibliography}
\end{document}